\documentclass[prb,twocolumn,showpacs,preprintnumbers,amsmath,amssymb]{revtex4}
\usepackage{graphicx}
\usepackage{dcolumn}
\usepackage{bm}
\usepackage{color}

\newcommand{\ur}{\ensuremath{\uparrow}}
\newcommand{\dr}{\ensuremath{\downarrow}}
\newcommand{\FMS}{Fe$_{3-x}$Mn$_x$Si}
\newcommand{\alat}{\ensuremath{a}}
\newcommand{\vc}{\vec}
\newcommand{\mub}{\ensuremath{\mu_{\rm B}}}
\newcommand{\EF}{\ensuremath{E_{\rm F}}}
\newcommand{\tg}{\ensuremath{t_{2g}}}
\newcommand{\eg}{\ensuremath{e_{g}}}
\newcommand{\tu}{\ensuremath{t_{1u}}}
\newcommand{\eu}{\ensuremath{e_{u}}}
\newcommand{\Tc}{\ensuremath{T_{\rm C}}}
\newcommand{\TR}{\ensuremath{T_{\rm R}}}
\newcommand{\FeA}{Fe$_{\rm A}$}
\newcommand{\FeB}{Fe$_{\rm B}$}
\newcommand{\FeC}{Fe$_{\rm C}$}
\newcommand{\FeAC}{Fe$_{\rm A,C}$}
\newcommand{\MnB}{Mn$_{\rm B}$}

\date{December 15, 2010}

\begin{document}

\title{Complex magnetic behavior and high spin polarization in
  Fe$_{3-x}$Mn$_x$Si alloys}

\author{Marjana Le\v{z}ai\'{c}}\email{M.Lezaic@fz-juelich.de}
\affiliation{Institut f\"ur Festk\"orperforschung, Institute for
  Advanced Simulation, Forschungszentrum J\"ulich and JARA, D-52425
  J\"ulich, Germany}
\author{Phivos Mavropoulos}\email{Ph.Mavropoulos@fz-juelich.de}
\affiliation{Institut f\"ur Festk\"orperforschung, Institute for
  Advanced Simulation, Forschungszentrum J\"ulich and JARA, D-52425
  J\"ulich, Germany}
\author{Stefan Bl\"ugel}
\affiliation{Institut f\"ur Festk\"orperforschung, Institute for
  Advanced Simulation, Forschungszentrum J\"ulich and JARA, D-52425
  J\"ulich, Germany}
\author{Hubert Ebert} \affiliation{Department
  Chemie/Physikalische Chemie, Ludwig-Maximillians-Universit\"at
  M\"unchen, Butenandtstr.~5-13, D-81377 M\"unchen, Germany}

\begin{abstract}
  Fe$_3$Si is a ferromagnetic material with possible applications in
  magnetic tunnel junctions. When doped with Mn, the material shows a
  complex magnetic behavior, as suggested by older experiments. We
  employed the Korringa-Kohn-Rostoker (KKR) Green function method within
  density-functional theory (DFT) in order to study the alloy
  Fe$_{3-x}$Mn$_x$Si, with $0\leq x\leq 1$. Chemical disorder is
  described within the coherent potential approximation (CPA). In
  agreement with experiment, we find that the Mn atoms align
  ferromagnetically to the Fe atoms, and that the magnetization and
  Curie temperature drop with increasing Mn-concentration $x$. The
  calculated spin polarization $P$ at the Fermi level varies strongly
  with $x$, from $P=-0.3$ at $x=0$ (ordered Fe$_3$Si) through $P=0$ at
  $x=0.28$, to $P=+1$ for $x>0.75$; {\it i.e.}, at high Mn concentrations
  the system is half-metallic. We discuss the origin of the trends of
  magnetic moments, exchange interactions, Curie temperature and the
  spin polarization.
\end{abstract}

\pacs{75.50.Bb, 71.20.Be, 71.70.Gm, 71.20.Lp}

%

\maketitle

\section{Introduction \label{sec:intro}}

Magnetic intermetallic alloys show very rich physics depending on
the degree of doping and chemical
disorder, which can therefore be used as ``control parameters'' allowing to tune
the electronic and magnetic structure for desired effects. 
In particular, physical properties that are fundamental for technological
applications in spintronics, such as the magnetization $M$, Curie
temperature \Tc, or spin polarization $P$ at the Fermi energy, vary
strongly with respect to these control parameters 

The alloy \FMS, belonging to the wider class of Fe$_{3-x}$TM$_x$Si
alloys with TM a transition-metal element,\cite{Niculescu83} is an
example of such dependence on the degree of doping.\cite{Yoon77} As
the Mn concentration increases, the magnetization of \FMS\ drops
continuously from 5~\mub\ to zero; its temperature-dependent magnetic
properties change from high-\Tc\ ($\approx$800~K) ferromagnetism,
through low-\Tc\ ferromagnetism with re-entrant behavior at 70~K, to
complex non-collinear magnetism; its calculated spin polarization
increases from $-30\%$ to the ideal, half-metallic
$+100\%$,\cite{Fujii95} and then drops again due to the non-collinear
behavior. These observations are not new, however, there is a recent
revival of the interest in \FMS\ due to potential applications in
magnetic tunnel junctions.\cite{Hayama09}

Motivated by this revival, we present here a theoretical study to
the electronic and magnetic properties of \FMS\ for $0<x<1$ based on
{\it ab initio} calculations. The choice of concentration range is
motivated by the specific site preference of Mn for $x<1$, so that the
resulting state is ferromagnetic as we see later; for $x>1$,
non-collinear magnetic configurations can occur. We provide an
interpretation of the magnetization drop as a function of
concentration in terms of wavefunction symmetry and hybridization,
together with the requirement for local charge neutrality. We further
propose that the increase in spin polarization up to the half-metallic
point is due to the same mechanisms that cause the magnetization
drop. Moreover, after extracting exchange interactions from the {\it
  ab initio} results, we calculate the Curie temperature using a Monte
Carlo approach, and are able to reproduce the drop of \Tc\ as a
function of Mn concentration. Finally, we discuss where our results do
not agree with experiment, and we propose a possible reason for the
disagreement; this is particularly the case for the re-entrant
behavior and the value of magnetization at high Mn concentrations.

The paper is structured as follows. In Sec.~\ref{sec:background} we
summarize the experimental and theoretical background on
\FMS. Sec.~\ref{sec:method} is devoted to the description of our
calculational approach. We continue with a presentation of our results
on the magnetization and spin polarization in Secs.~\ref{sec:moments}
and \ref{sec:dos} and of the Curie temperature in
Sec.~\ref{sec:tc}. In Sec.~\ref{sec:limitations} we discuss the
limitations of our approach, their consequences, and possibilities for
a more accurate description. Our conclusions are summarized in
Sec.~\ref{sec:conclusion}.

\section{Experimental and theoretical
  background \label{sec:background}}

\subsection{Experimental}

Quite a few experimental studies have been done on the magnetic
properties of \FMS, revealing a highly complex magnetic behavior
dependent on the Mn concentration $x$. Here we recall the main results
of these experiments, with emphasis on the concentration range $0<x<1$
which interests us in the present work. At high concentrations, one
reaches the Fe-doped Mn$_3$Si compound, which exhibits more
complicated properties; {\it e.g.}, Mn$_3$Si is an incommensurate
antiferromagnet with a N\'eel temperature of about 25~K, while Fe$_3$Si
is a ferromagnet.

\begin{figure}
\begin{center}
\includegraphics[width=8cm]{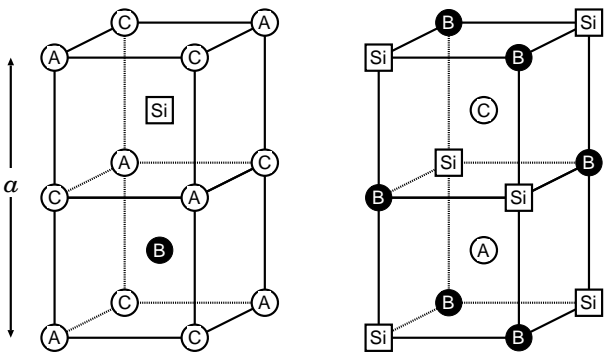}
\end{center}
\caption{Geometrical structure of Fe$_3$Si and \FMS. The A and C sites
  are occupied by Fe, while the B site is occupied either by Fe or by
  Mn. Square symbols show the positions of the Si atoms. In the left
  panel we show the octahedral coordination of the B-site to A and C
  Fe neighbors, while in the right panel we show the tetrahedral
  environment of the A and C sites.\label{fig:geometry}}
\end{figure}

{\bf Structure and site preference.}  Fe$_3$Si crystallizes in the
D0$_3$ structure consisting of a fcc lattice with four basis
atoms (see Fig.~\ref{fig:geometry}). These are placed at $(0,0,0)$\alat\ (Si atom),
$(\frac{1}{4},\frac{1}{4},\frac{1}{4})$\alat\ (\FeA-atom),
$(\frac{1}{2},\frac{1}{2},\frac{1}{2})$\alat\ (\FeB-atom),
$(\frac{3}{4},\frac{3}{4},\frac{3}{4})$\alat\ (\FeC-atom), where
$\alat$ is the lattice constant. The \FeA\ and \FeC\ atoms are
tetrahedrally coordinated to four \FeB\ and four Si atoms and exhibit
equivalent electronic ground-state properties due to symmetry. The
\FeB\ atoms are octahedrally coordinated to eight \FeAC\ atoms; their
electronic properties therefore resemble somewhat bcc Fe, as we will
discuss.

When Mn is doped into Fe$_3$Si, it substitutes \FeB\ atoms, as is
found by experiment. This appears to be part of a general trend found
experimentally\cite{Burch74,Pickart75,Niculescu83} and modeled
theoretically\cite{Swintendick76} in which transition-element atoms
which are to the left of Fe in the periodic table prefer to reside at
the B-site when doped into Fe$_3$Si, while transition-element atoms
which are to the right of Fe prefer to substitute the Fe atoms at the
A and C sites.

Yoon and Booth\cite{Yoon74,Yoon77} report that, in the range
$0<x<0.75$, the Mn atoms substitute the B-site Fe, as verified by
hyperfine field measurements of Niculescu {\it et
  al.}.\cite{Niculescu76,Niculescu83} Above $x=0.75$, an increasing
fraction of Mn atoms tends to reside at the A,C sites with equal
probability. At $x=1$, this ``swap'' fraction is 12\% according to
Ziebeck and Webster\cite{Ziebeck76} and 15\% according to Yoon and
Booth\cite{Yoon77}; {\it i.e.}, even Fe$_2$MnSi contains a small
amount of chemical disorder. \FMS\ has thus the L2$_1$, or
full-Heusler, structure.  Similar results are found in more recent
hyperfine-field experiments.\cite{Mahmood04}

The measured lattice parameter changes
linearly and only slightly as a function of the concentration $x$ (see
Ref.~\onlinecite{Niculescu76}) from 5.653~\AA\ at $x=0$ to 5.663~\AA\
at $x=1$, {\it i.e.,} by about 0.2\%. The linear change continues for
higher $x$.

{\bf Magnetic moments.}  As was shown by Booth {\it et
  al.}\cite{Booth74} and Yoon and Booth,\cite{Yoon74,Yoon77} the
saturation magnetization $M$ drops linearly from about 4.8~\mub\ to
2.6~\mub\ per formula unit in the range $0<x<0.75$.  The local
magnetic moments depend strongly on the site. The B-site has a high
moment of about $2.3\pm 0.3\ \mub$ between $0\leq x<0.75$, which then
drops gradually and vanishes at $x=1.75$; at the A,C sites (containing
only Fe for low $x$), the moment decreases from 1.4~\mub\ to 0.3~\mub\
as $x$ increases from 0 to 0.75. The net result is the aforementioned
drop of the total magnetic moment.

{\bf Anomalous temperature dependence of the magnetization.} In the
concentration range $0.75<x<1.75$ a re-entrant behavior of the
magnetization curve $M(T)$ is found:\cite{Yoon74,Yoon77} for
temperatures from $T=0$~K up to the re-entry temperature $\TR\approx
70$~K, $M(T)$ is \emph{increasing}, while for $T>\TR$ $M(T)$ is
\emph{decreasing}, as expected for a usual ferromagnet, up to the
Curie temperature $\Tc$. Thus, two values of the saturation
magnetization can be defined, one ($M_0$) as the actual measured value
$M(T=0)$, and another ($M_{\mathrm{extr}}$) as the extrapolated value
of $M(T)$ from data taken for $T>\TR$; evidently,
$M_{\mathrm{extr}}>M_0$. For $x=0.75$, in fact, where this effect just
starts to appear, the two values are close
($M_0\approx0.95M_{\mathrm{extr}}$), and application of a magnetic
field of $14$~kOe can further saturate the sample so as to reach the
value $M_{\mathrm{extr}}$. The anomalous temperature dependence of $M$
was also found by Ziebeck and Webster~\cite{Ziebeck76} for the
Fe$_2$MnSi alloy, as well as by Nagano and co-workers\cite{Nagano95}
and by Ersez et al.\cite{Ersez96} who analyzed the effect via neutron
scattering. However, in a recent work where the samples were grown by
molecular beam epitaxy,\cite{Hayama09} this anomalous behavior was not
observed (up to $x=1.4$).

{\bf Curie temperature.} The Curie temperature drops as a function of
the Mn concentration $x$, from about $\Tc\approx 800$~K at $x=0$ to
$\Tc=0$~K at $x\approx 1.75$; {\it i.e.}, for $x>1.75$ no saturation
magnetization is found, while the sample is at a complex non-collinear
magnetic state. At $x=0.75$, where the re-entrant magnetic behavior sets
on, $\Tc=375$~K and $\TR\approx 40$~K, while at $x=1$, $\Tc\approx
220$~K and $\TR\approx 70$~K.\cite{Yoon77}

{\bf Thermodynamic properties.} The anomalous behavior at $\TR$ also
shows in measurements of thermodynamic quantities.  In particular,
Smith {\it et al.}\cite{Smith80} have measured a specific heat anomaly
at $\TR$, most pronounced for the $x=1$ compounds. Furthermore, Miles
{\it et al.}\cite{Miles88} report a sharp peak of the thermal
expansion coefficient at $\TR=60$~K for the Fe$_2$MnSi alloy ({\it i.e.},
$x=1$), while they find no such pronounced behavior at $\Tc$.

{\bf Heat treatment.} As reported in Ref.~\onlinecite{Yoon77}, the
sample preparation included a 24-hour heating at 830$^{\circ}$C and
water quenching; a different heat treatment (21 days at 550$^{\circ}$C
and slow cooling) for samples with $0.95<x<1.25$ appeared to increase
$M_0$, bringing it closer to $M_{\mathrm{extr}}$, but had no effect on
the values of $\TR$ or $\Tc$. Ziebeck and Webster\cite{Ziebeck76} and
Smith {\it et al.}\cite{Smith80} also used an annealing treatment at
over 800$^{\circ}$C for 24 hours water quenching. Miles {\it et
  al.}\cite{Miles88} used two samples, one quenched from
800$^{\circ}$C and one slowly cooled, with no change in the expansion
coefficient and its anomaly at $\TR$.

\subsection{Theoretical}

Only few theoretical results exist on the electronic structure of
\FMS. Mohn and Supanetz\cite{Mohn98} employed an augmented spherical
wave method and the local spin density approximation (LSDA) to
density-functional theory to examine non-collinear states in the
ordered alloys Fe$_3$Si, Fe$_2$MnSi (with the Mn atom at the B-site),
FeMn$_2$Si (with one Mn atom at the B-site and one at the A-site), and
Mn$_3$Si. For Fe$_3$Si they found a ferromagnetic ground state, while
non-collinear ground states were found for all other compounds. In
particular for Fe$_2$MnSi, which is of interest here, Mohn and
Supanetz\cite{Mohn98} found a local energy minimum at the
ferromagnetic state, with a lower minimum for a spin-spiral along the
[111] axis, at a $\mathbf{q}$-vector of
$\frac{2\pi}{a}(0.45,0.45,0.45)$.  According to their calculation, the
ground state was an antiferromagnetic state, with the Mn moments
alternating along the [111] axis, while the moment direction was
canted off the [111] direction by about 60$^{\circ}$. The energy
difference between the non-collinear ground state and the
ferromagnetic state in Fe$_2$MnSi was reported to be around 0.8~mRyd
(10.9~meV). A canting of the magnetic moments below $\TR$ was also
assumed by Yoon and Booth\cite{Yoon74,Yoon77} in order to explain the
neutron scattering data.

More recently, {\it ab initio} calculations were presented\cite{Go07}
on \FMS\ for $0\leq x\leq 0.5$ using the supercell method. Mn was
considered to occupy the B site. It was found that the spin moments of
Fe at the A and C sites are reduced in the presence of Mn nearest
neighbors, which induces a redistribution of the \FeAC\ states; a drop
of the total moment with increasing Mn concentration was observed and
attributed to the \FeAC\ moments. Furthermore, {\it ab initio}
calculations on  \FMS\ and Fe$_{3-x}$MnSi$_x$ alloys were presented in
Ref.~\onlinecite{Hamad10}; the calculations here agree with the
previous results that the magnetic moments of the A and C sites drop
as a function of Mn concentration, while it is found that the B-site
atomic moments increase. The Fe moments appear to be higher than the
Mn moments, so that the B-site average moment does not change much.

Further theoretical work appears in parallel with
experiments. Szyma\'nski~{\it et al.}\cite{Szymanski91} examined the
spin dynamics of \FMS\ using neutron scattering at room temperature
and at liquid nitrogen temperature, and fitted their results to
effective interatomic exchange integrals which enter a Heisenberg
Hamiltonian. The fitted values of the exchange constants depend on the
number of neighbors considered; the nearest-neighbor exchange for
Fe$_3$Si ranges between 10 and 20~meV. Brown and
co-workers\cite{Brown85} analyze the behavior of the magnetic moments
based on symmetry arguments and on a model by
Swintendick\cite{Swintendick76} and conclude that, as the Mn
concentration increases, the reduction of the exchange splitting leads
to the drop of the \FeAC\ moment.

\section{Method of calculation, models and cutoff parameters \label{sec:method}}

Our electronic-structure calculations are based on density-functional
theory within the generalized gradient approximation (GGA)\cite{PW91}
to account for exchange and correlation effects. The local spin
density approximation\cite{Vosko} was also used for comparison and
proved to be inadequate for the prediction of the correct magnetic
ground state at low and intermediate Mn concentrations (see
Sec.~\ref{sec:moments}). Calculations were also performed within the
``single-shot GGA'',\cite{Asato99} where, using the self-consistent
LSDA spin density, $\boldsymbol{\rho}_{\mathrm{LSDA}}$, the total
energy is calculated within the GGA functional
$E_{\mathrm{GGA}}[\boldsymbol{\rho}]$; {\it i.e.\ } one calculates
$E_{\mathrm{GGA}}[\boldsymbol{\rho}_{\mathrm{LSDA}}]$. This approach
is based on the idea that $\boldsymbol{\rho}_{\mathrm{LSDA}}$, as a
trial density, is not too far from the self-consistent GGA density
$\boldsymbol{\rho}_{\mathrm{GGA}}$, so that, due to the variational
character of the energy functional,
$E_{\mathrm{GGA}}[\boldsymbol{\rho}_{\mathrm{LSDA}}]\approx
E_{\mathrm{GGA}}[\boldsymbol{\rho}_{\mathrm{GGA}}]$. The single-shot
GGA is known, for instance, to correct the LSDA overbinding, giving an
improved equilibrium lattice parameter, very close to the one
predicted by the GGA.\cite{Asato99}

The Kohn-Sham equations are solved in most cases within the
full-potential Korringa-Kohn-Rostoker (KKR) Green function
method\cite{SPRTBKKR} with exact treatment of the atomic cell
shapes,\cite{Stefanou91} using an angular momentum cutoff of $l_{\rm
  max}=3$ and an integration mesh of $30\times 30\times 30$ points in
the full Brillouin zone. The substitutional disorder was described
within the coherent potential approximation (CPA). For the calculation
of magnetically non-collinear states and static magnon spectra we
employed the full-potential linearized augmented plane
wave~\cite{FLAPW1,FLAPW2,FLAPW3} (FLAPW) method as implemented in the
{\tt FLEUR} code,\cite{FLEUR} using a plane-wave cutoff of $k_{\rm
  max}=4\ {\rm a.u.}^{-1}$, an angular momentum cutoff of $l_{\rm
  max}=8$, muffin-tin radii of 1.19 \AA\ for Fe and Mn and 1.222 \AA\
for Si, and an $17\times 17\times 17$ mesh in the full Brillouin
zone. The FLAPW code was also used to cross-check the KKR results in
some cases. Relativistic effects were taken into consideration within
the scalar relativistic approximation, whereas spin-orbit coupling was
not accounted for.

Since the lattice parameter varies only slightly\cite{Niculescu76} in
the range $0\leq x\leq 1$ we used the experimental value at $x=0$,
$\alat=5.653$~\AA, in all calculations.

Magnetic excitations are modeled within a classical Heisenberg
Hamiltonian,
\begin{eqnarray}
  H &=& -\sum_{ij} J_{ij} \hat{e}_{i} \cdot \hat{e}_{j},\\
  \label{eq:heisenberg}
  &=& -\sum_{ij} \tilde{J}_{ij} \vec{M}_{i} \cdot \vec{M}_{j}
  \label{eq:heisenbergB}
\end{eqnarray}
where $ \hat{e}_{i}$ and $\hat{e}_{j}$ are unit vectors along the
directions of the spin moments, $\vec{M}_{i}$ and $\vec{M}_{j}$, at
sites $i$ and $j$, while the exchange pair-interaction constants
$J_{ij}$ reflect the energy cost for the mutual tilting of the
moments. It is sometimes convenient for the discussion to use the form
(\ref{eq:heisenbergB}) with $\tilde{J}_{ij}=J_{ij} /
(M_{i}M_{j})$. The constants $J_{ij}$ were extracted from the
spin-dependent KKR structural Green functions 
$G_{ij}^{\ur(\dr)}(E)$ and $t$-matrices $t_i^{\ur(\dr)}(E)$ by virtue of the
Liechtenstein formula\cite{Liechtenstein87}
\begin{equation}
  J_{ij} = \frac{1}{4\pi} \mathrm{Im}\,
  \mathrm{Tr} \int^{\EF}  G^{\ur}_{ij}\, (t^{\ur}_j - t^{\dr}_j )\,
  G^{\dr}_{ji}\, (t^{\ur}_i - t^{\dr}_i )\, dE.
  \label{eq:liechtenstein}
\end{equation}
Here, $G_{ij}^{\ur(\dr)r}(E)$ and $t_i^{\ur(\dr)}(E)$ are matrices in
angular momentum space and $\mathrm{Tr}$ denotes a trace over the
angular momentum indices ($lm$).

Having calculated the exchange constants, the Curie temperature of the
compounds was calculated within a Monte Carlo method. For this purpose,
exchange constants of atom pairs $(i,j)$ with distance up to $2.18$
lattice constants were used; the simulation supercells included 1536
magnetic atoms (512 unit cells). The Curie temperature was identified
through the characteristic peak of the calculated susceptibility.  The
method, either in combination with Monte Carlo simulations or with the
random phase approximation for the solution of the Heisenberg model,
has proven useful for the calculation of the exchange constants and
Curie temperature with a 10-15\% accuracy in several cases, including
elemental ferromagnets and intermetallic
alloys.\cite{Sasioglu05,Turek06,Lezaic07}

\section{Ground state magnetic moments and
  configuration \label{sec:moments}}

\begin{figure}
  \includegraphics[width=8cm]{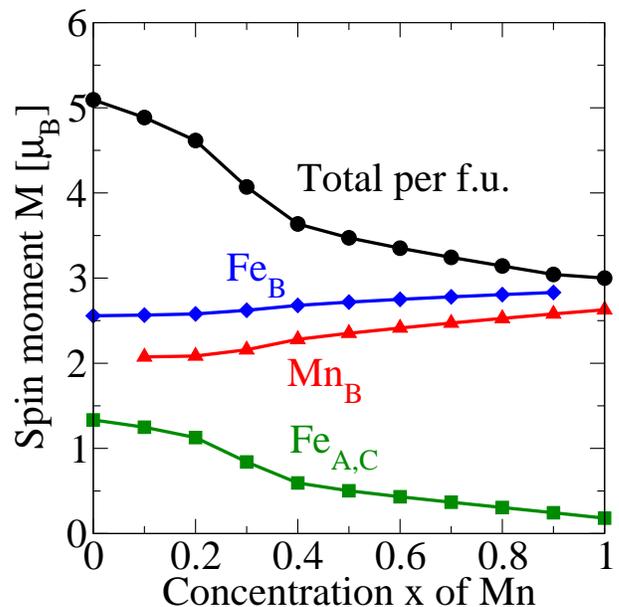}
  \caption{(color online) Calculated magnetic moments per atom and
    total magnetic moment in the unit cell as a function of Mn
    concentration in \FMS.\label{fig:moments}}
\end{figure}

The calculated magnetic moments (per atom and total) as a function of
concentration are depicted in Fig.~\ref{fig:moments}. Here the Mn
atoms were assumed to reside at the B site for all $x$. Evidently, the
B-site atoms (Fe and Mn alike) are generally in a high-spin state,
while the Fe atoms at the A and C sites are in a low-spin state. A
small, monotonic increase of the B-site atomic moment is observed as a
function of the Mn concentration: the \FeB\ moment ranges between 2.6
and 2.9~\mub, while the Mn moment ranges between 2.1 and 2.6~\mub. In
strong contrast, the \FeAC\ moments drop significantly as the Mn
content increases, from 1.3~\mub\ for pure Fe$_3$Si to 0.2~\mub\ for
the ordered Fe$_2$MnSi alloy. This drop of the \FeAC\ moment
causes the decrease in the total magnetic moment per unit cell.

This trend is in agreement with experimental
findings\cite{Booth74,Yoon74,Yoon77} that the average moment at the
B-site is high and remains more or less unaffected by Mn-doping, while
the A and C-site moments drop significantly. (A deviation from the
experimental result is found for $x\geq 0.8$, where there is
experimental evidence for re-entrant behavior and reordering of
spins.) The trend can be understood by an analysis of the density of
states and an understanding of the different wavefunction
hybridization of Mn$_{\mathrm{B}}$-\FeAC\ and \FeB-\FeAC\ atoms; we
defer the discussion to section \ref{sec:dos}.

\begin{figure}
  \begin{center}
    \includegraphics[width=8cm]{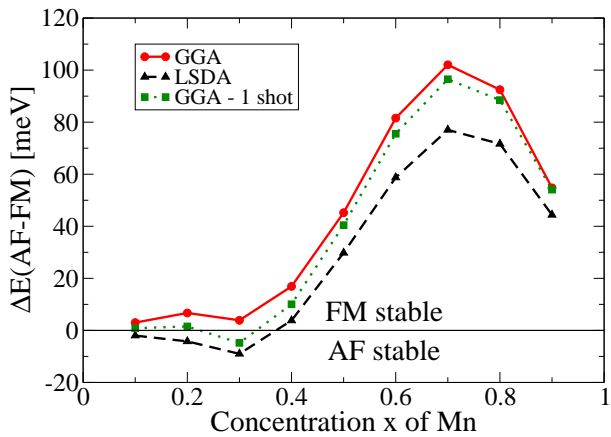}
    \caption{(color online) Results of total-energy calculations on
      the magnetic state of \FMS. The energy difference $\Delta
      E=E(\mathrm{AF})-E(\mathrm{FM})$ between the antiferromagnetic
      and the ferromagnetic alignment of Mn atoms with respect to the
      \FeB\ atoms is shown ($\Delta E$ represents here values per unit
      cell). The GGA predicts the correct ferromagnetic state, while
      the LSDA yields an antiferromagnetic orientation of the Mn
      moments with respect to the Fe$_3$Si matrix at low
      concentrations, which turns to ferromagnetic at high
      concentrations. A single-shot GGA calculation improves the LSDA
      result somewhat, but not quite.\label{fig:energy}}
  \end{center}
\end{figure}

In the calculation, the Mn dopants can be chosen to align
ferromagnetically or antiferromagnetically to the Fe$_3$Si
matrix. Total-energy calculations are then needed to identify the
correct ground state, which experimentally is found to be
ferromagnetic,\cite{Booth74} at least for concentrations $x<0.75$. The
calculations within the GGA show that the ferromagnetic state is
stable for all Mn concentrations $0<x\leq 1$ (see
Fig.~\ref{fig:energy}). However, the result within the LSDA is that
the antiferromagnetic state is more stable than the ferromagnetic one
for $x\lesssim 0.35$, in clear disagreement with
experiment;\cite{Booth74} actually, for $0.2 \lesssim x \lesssim 0.8$,
the LSDA lowest energy is found in a disordered local moment state of
the form Fe$_{3-x}$Mn$^\downarrow_{x-y}$Mn$^\uparrow_{y}$Si, as the Mn
atoms progressively change the moment orientation from
antiferromagnetic (Mn$^\downarrow$) to ferromagnetic
(Mn$^\uparrow$). This indicates a possible non-collinear LSDA ground
state (our calculations for the disordered alloys were always
magnetically collinear). The dispute between the LSDA and GGA results
at low concentrations was cross-checked and verified by a calculation
within the FLAPW method, where a low Mn concentration of $x=0.125$ was
approximated by construction of a large supercell (the same lattice
parameter was used in LSDA and GGA calculations, see
Sec.~\ref{sec:method}).

The failure of the LSDA to predict the correct magnetic ground state
can be attributed to the exchange and correlation part of the total
energy, rather than the single-particle energies. We arrive at this
conclusion for two reasons. Firstly, the GGA density of states is very
similar to the LSDA density of states (when both are calculated in the
ferromagnetic configuration). Secondly, we attested our suggestion by
using the single-shot GGA (described in Sec.~\ref{sec:method}), which
changes only the exchange-correlation part of the total energy, 
while
retaining the LSDA single-particle energies. The total-energy results
are then improved significantly, although not entirely, towards the
correct magnetic ground state as can be seen in
Fig.~\ref{fig:energy}. We note that there is no general rule favoring
the GGA over the LSDA as far as the magnetic properties are
concerned. For example, long-wavelength spin-wave spectra calculated
within the adiabatic approximation agree rather well with experiment
if the LSDA is used (see, {\it e.g.}, Pajda {\it et al}.\cite{Pajda01} for
Fe, Co and Ni, or Buczek {\it et al}.\cite{Buczek09} for intermetallic
alloys). There are even reported
cases, such as Fe$_3$Al,\cite{Lechermann02} where GGA gives the wrong
crystal structure, while LSDA corrects the structure as well as the
magnetic moments.

\section{Density of states, magnetic moments, and half-metallic
  behavior \label{sec:dos}}

\subsection{Trends of DOS with concentration; magnetic moments}

\begin{figure*}
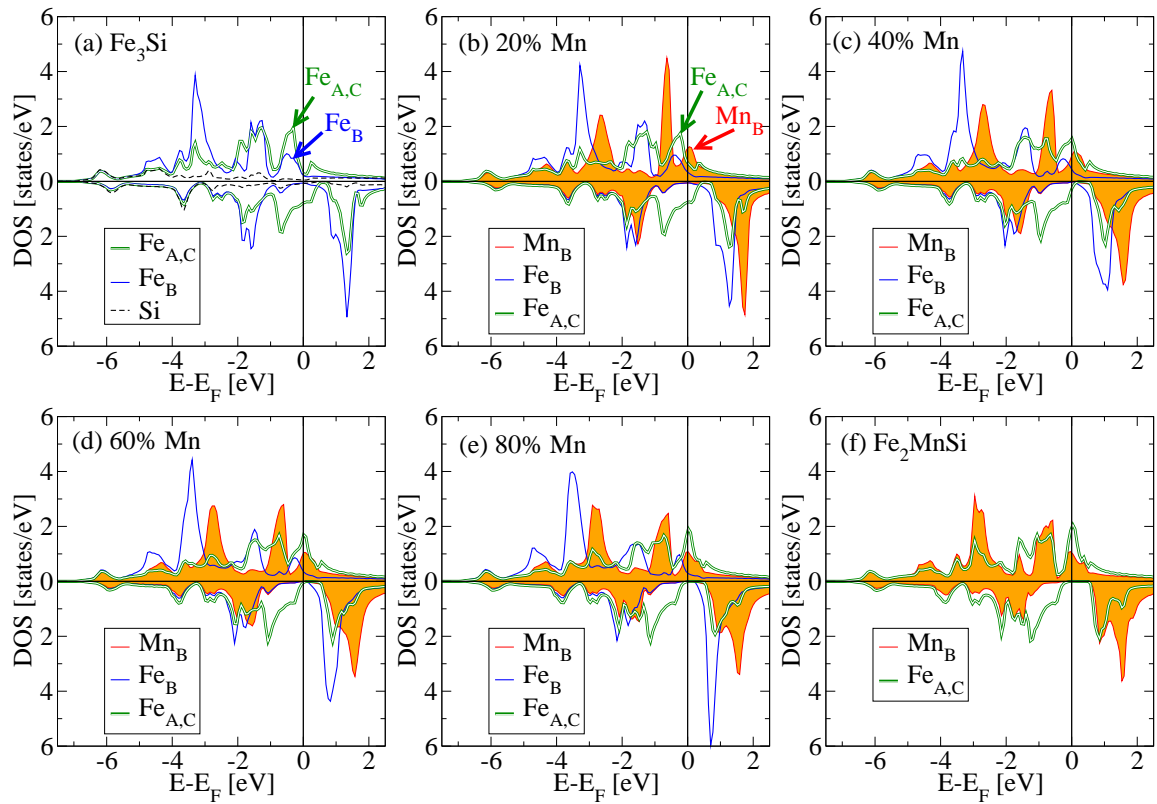

  \begin{center}
    \includegraphics[width=5cm]{dosFe3Si_GGA.eps}
    \includegraphics[width=5cm]{dos20F_GGA.eps}
    \includegraphics[width=5cm]{dos40F_GGA.eps}
    \includegraphics[width=5cm]{dos60F_GGA.eps}
    \includegraphics[width=5cm]{dos80F_GGA.eps}
    \includegraphics[width=5cm]{dosFe2MnSi_GGA.eps}
  \end{center}
  \caption{(color online) Spin- and atom-resolved density of states of
    \FMS\ alloys for Mn concentrations $0\leq x \leq 1$. In all cases,
    the Mn atom taken to reside at the B-site. Up and down arrows
    indicate the majority and minority-spin, respectively. In (a) and
    (b), arrows indicate the positions of the DOS peaks near \EF\
    which are responsible for the moment trends, as discussed in the
    text. The calculations were done within the KKR-CPA. In each plot,
    the upper panel corresponds to spin-up (majority-spin) DOS and the
    lower panel (with inverted ordinate) to spin-down (minority-spin)
    DOS. The orange-shaded area under the red line corresponds to Mn,
    the full blue line to \FeB\ and the double green line to
    \FeAC. \label{fig:dos}}
\end{figure*}

The density of states (DOS) of \FMS\ at various concentrations $x$ is
depicted in Fig.~\ref{fig:dos}. The gross features have been analyzed
in the past in studies of full Heusler alloys.\cite{Galanakis02} Most
important points specifically for our discussion are: (i) The
hybridization of $d$-states of \tg\ character of the B-site atom (Fe
or Mn) with \tg\ states of the A and C-site Fe atoms. (ii) The
strong-ferromagnet character of the B-atoms as opposed to the
weak-ferromagnet character of the (A,C) atoms.  (iii) The progressive
shift, for charge-neutrality reasons, of the majority-spin states at
the B-site as the Mn concentration is increased, dragging with them
the (A,C)-site \tg\ states due to hybridization and affecting the
\FeAC\ moment and DOS. We now discuss these points in detail; the
trends of the magnetization and the appearance of half-metallicity
are directly connected to this behavior and also discussed in
this section.

\begin{figure}
  \begin{center}
    \includegraphics[width=8cm]{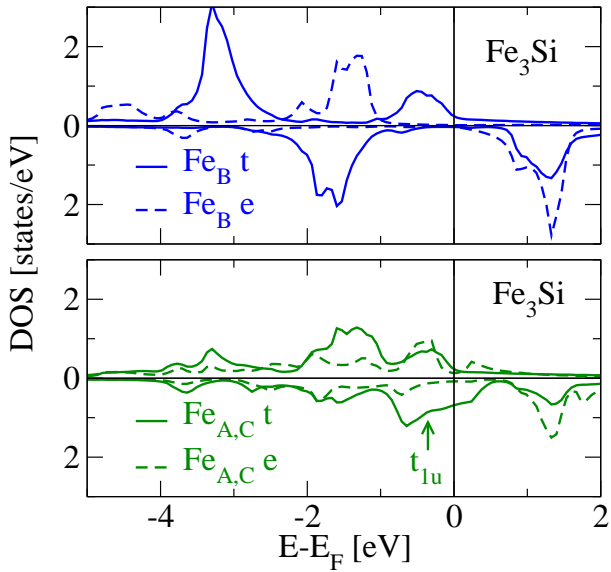}
  \end{center}
  \caption{(color online) Symmetry-resolved local density of states of
    Fe$_3$Si at the \FeB\ (top) and \FeAC\ (bottom) atoms. In each plot,
    the upper panel corresponds to spin-up and the lower panel to
    spin-down DOS.\label{fig:lmdos}}
\end{figure}

(i) The $d$-states of \FeB\ and Mn are split by the octahedral
environment into two irreducible representations: the \tg, including
the $d_{xy}$, $d_{yz}$, and $d_{xz}$ orbitals, and the \eg, including
the $d_{x^2-y^2}$ and $d_{z^2}$ orbitals ($x$, $y$, and $z$ are
implied to be directions along the cubic crystal axes). The \FeAC\
$d$-states are also split by the tetrahedral environment. However, it
is also important that $d$ orbitals of \FeAC\ atoms interact with each
other, although the \FeAC\ atoms are second-nearest neighbors. This
interaction produces bonding-antibonding splittings between \tg\ and
\tu-type of states, and also between \eg\ and \eu-type of
states.\cite{Galanakis02}

Hybridization between $d$-states of B-site atoms with \FeAC\ atoms is
allowed only among the states of \tg\ or the ones of \eg\ character;
by symmetry, the \tu\ and \eu\ states remain oblivious to the
$d$-states of their B-site neighbors.\cite{Galanakis02} The
symmetry-decomposed DOS is shown in Fig.~\ref{fig:lmdos} for Fe$_3$Si,
and is completely analogous in \FMS, with the peaks appropriately
shifted as discussed below.

\begin{figure}
  \begin{center}
    \includegraphics[width=8cm]{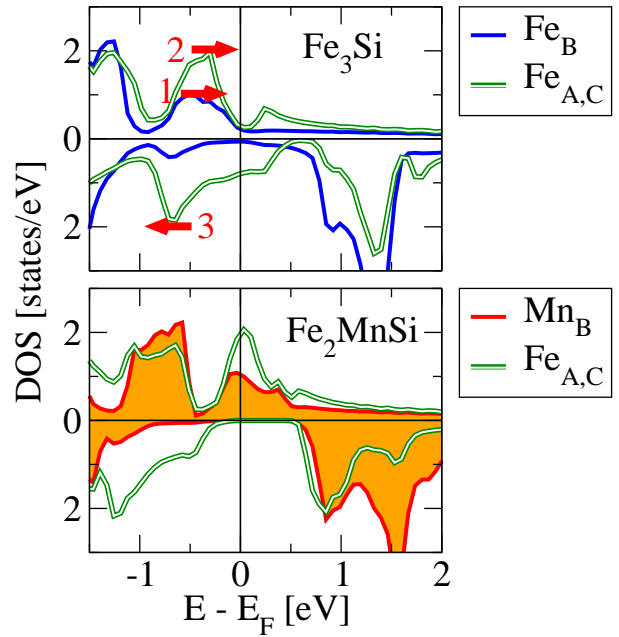}
  \end{center}
  \caption{(color online) Spin- and atom-resolved density of states of
    Fe$_3$Si (top) and Fe$_2$MnSi (bottom) around \EF. Arrows indicate
    the shift of the peaks upon substitution of Fe by Mn that lead to
    the reduction of the moment and half-metallic behavior. 1. The
    B-site spin-up peak shifts higher in order to dispose of one
    electron. 2. The spin-up \FeAC\ peaks are dragged along due to
    hybridization, disposing of spin-up charge also at the \FeAC\
    atoms. 3. The spin-down peak at \FeAC\ must shift lower in order
    to maintain local charge neutrality. The result is a reduction of
    the magnetic moment and an increase of the spin polarization at
    \EF. In each plot, the upper panel corresponds to spin-up and the
    lower panel to spin-down DOS. The orange-shaded area under the red
    line corresponds to Mn, the full blue line to \FeB\ and the double
    green line to \FeAC.\label{fig:mechanism}}
\end{figure}

(ii) On the one hand, the local DOS of the B-site atoms (either Fe or
Mn) has in all cases the characteristics of the DOS of a \emph{strong
  ferromagnet}. Namely, the B-site local DOS of one spin direction
(here spin down) is very low at \EF\ and in a region around \EF, as is
evident by inspection of Fig.~\ref{fig:dos}. Consequently,
\emph{on-site} transfer between spin-up and spin-down charge is
energetically expensive, because a strong shift of $d$-bands is
involved. This, together with the requirement of local charge
neutrality in a metallic system, stabilizes the local atomic moment of
Fe or Mn at the B-site against perturbations (such as change of
concentration or lattice parameter). Therefore there is only weak
dependence of the Fe and Mn moment on concentration, as seen in
Fig.~\ref{fig:moments}. The strong-ferromagnet behavior of the B-site
is favored by its octahedral bcc-like coordination, which results in
large bonding-antibonding splittings via \tg-\tg\ and
\eg-\eg-hybridization of the B-site atoms with \FeAC\ neighbors. The
line-shape of the local DOS at the B-site is reminiscent of bcc iron.

On the other hand, the local DOS of the \FeAC\ atoms has the
characteristics of a \emph{weak ferromagnet}, {\it i.e.}, $d$-states of both
spin directions are present at and around \EF. This allows for
energetically cheap transfer between spin-up density and spin-down
density, and therefore relatively easy change of moment. Note from
Fig.~\ref{fig:moments} that the drop of the \FeAC-moment is strongest
for $0.2<x<0.4$, at the same concentration range when the \FeAC-DOS at
\EF\ is large for both spins in Fig.~\ref{fig:dos}, {\it i.e.}, when the
weak-ferromagnet character is most evident; before $x=0.2$ there is
a DOS valley at \EF\ for spin-up, and after $x=0.4$ there is a valley
for spin-down. The weak-ferromagnet behavior of \FeAC\ is favored by
the \tu\ and \eu\ states forming the spin-down peak of \FeAC\ around
\EF\ in Fe$_3$Si.

(iii) Now we consider the consequences of observations (i) and (ii) (see
Figure~\ref{fig:mechanism}). We first focus on the B-site in
\FMS. Since a metal must show approximate local charge neutrality, the
Mn states must be appropriately shifted with respect to those of Fe so
that one less electron is accommodated by Mn. This is achieved by a
shift of the spin-up \tg\ and \eg\ peaks, which are just under \EF\
for the \FeB\ atom, so that they fall at \EF\ for the Mn atom (as
indicated by arrows in Fig.~\ref{fig:dos}(a,b)). Now, there is a peak
of \tg\ character (also indicated by arrows in
Fig.~\ref{fig:dos}(a,b)) in the \FeAC\ DOS, associated by
hybridization to the aforementioned B-site peak. This, at low Mn
concentrations, is associated more to the \FeB\ DOS, while at high Mn
concentrations it is associated more to the Mn DOS. At intermediate
concentrations, as the Mn content increases, the \FeAC\ peak is
dragged to higher energies and starts crossing the Fermi level,
depriving the \FeAC\ atoms of spin-up charge. This is readily
compensated (to maintain local charge neutrality) by a shift of
spin-down \FeAC\ \tu\ states from \EF\ to slightly below \EF, gaining
spin-down charge. The net effect is a reduction of the \FeAC\ spin
moment, accompanied by the appearance of a spin-down gap at \EF, {\it i.e.},
the signature of a half-metallic behavior.

Thus we see that the \tg-\tg\ hybridization between states at the B
and A,C sites, together with the requirement of local charge
neutrality, leads to the drop of \FeAC\ moment as the Mn concentration
increases.

In spite of the calculated drop of the magnetization per formula unit,
the values are still too high compared to experiment,\cite{Yoon77}
although they are in agreement with previous
calculations.\cite{Fujii95,Galanakis02,Go07,Hamad10} For Fe$_2$MnSi,
the calculated value is 3~\mub, while the experiment gives $M_0\approx
1.5~\mub$ and $M_{\rm extr}\approx~2.3\mub$. We propose an explanation
of this discrepancy in Sec.~\ref{sec:limitations}.

\subsection{Spin polarization and transition to half-metallic
  behavior}

\begin{figure}
  \begin{center}
    \includegraphics[width=6cm]{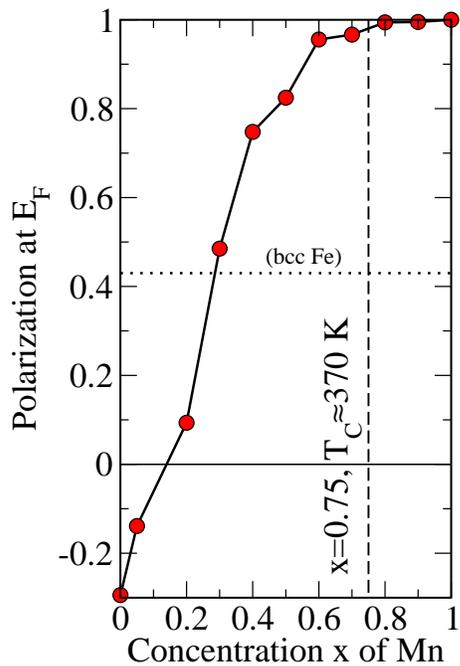}
  \end{center}
  \caption{Calculated spin polarization $P$ at \EF\ as
    a function of Mn concentration $x$ in \FMS. For $x\ge 0.7$ the
    alloy is half-metallic. The polarization of bcc Fe (dotted line)
    is also shown for comparison.\label{fig:pol}}
\end{figure}

The spin polarization $P$ at \EF\ is defined as
\begin{equation}
  P=\frac{n_{\ur}(\EF)-n_{\dr}(\EF)}{n_{\ur}(\EF)+n_{\dr}(\EF)}
  \label{eq:polarization}
\end{equation} 
with $n_{\ur}(\EF)$ and $n_{\dr}(\EF)$, respectively, the spin-up and
spin-down DOS at \EF. The mechanism described in the previous
subsection, involving a shift of the peaks around \EF, leads to a
drastic change of $P$ as a function of the Mn concentration. This is
visualized in Fig.~\ref{fig:pol}. At low $x$, $P$ is negative,
approximately $-0.3$ (the negative sign means that the DOS at \EF\ is
dominated by minority-spin carriers). At higher concentrations $P$
crosses zero and reaches high positive values, as the spin-up DOS peak
shifts to \EF, while the spin-down peak is retracted below \EF. At
around $x=0.75$ the spin polarization reaches the highest possible
value of $P=1$, and the alloy becomes half-metallic. At this point the
(experimentally found) Curie temperature is about 370~K. A
half-metallic behavior of ordered Fe$_2$MnSi was also found in
previous calculations.\cite{Fujii95,Galanakis02,Go07,Hamad10}

This is a rare occasion in which a continuous change of a material
parameter --- here the Mn concentration --- results in a continuous
change of the spin polarization over such an extended range. Assuming
that this effect is present in experiment, it could be efficiently
used to compare to each other various experimental methods of probing
the spin polarization (such as spin-polarized photoemission
spectroscopy, positron annihilation, Andreev reflection, or tunneling
magnetoresistance). It should be noted that the half-metallic behavior
is already present at $x=0.75$, {\it i.e.}, before the start of the
re-entrant behavior.

\section{Exchange interactions and Curie temperature\label{sec:tc}}

\begin{figure}
  \begin{center}
    \includegraphics[width=8cm]{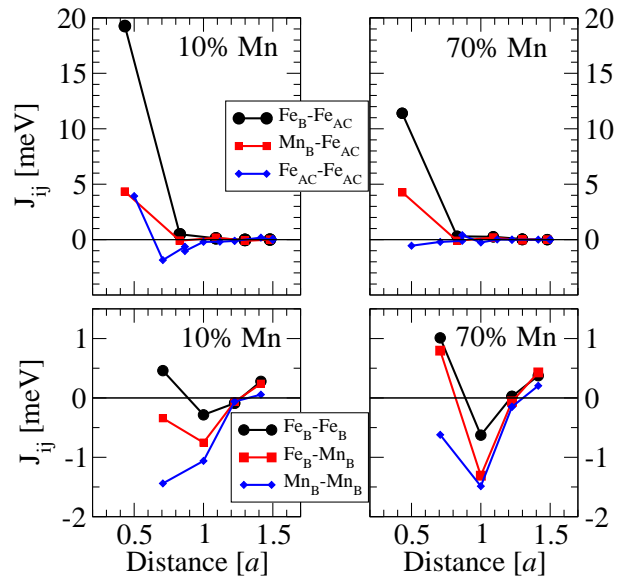}
    \caption{(color online) Pair exchange constants as a function of
      the inter-atomic distance. Left: 10\% Mn concentration. Right:
      70\% Mn concentration. Note the different energy scale in the
      upper and lower panels. The lines are guides to the eye.\label{fig:Jij}}
  \end{center}

\end{figure}

The exchange pair-interaction constants $J_{ij}$ were calculated for
several Mn concentrations as described in Sec.~\ref{sec:method}.  They
are plotted as a function of distance in Fig.~\ref{fig:Jij}. In all
cases, we find that the dominant contribution comes from the
first-neighbor interaction $J_1$ between the site-B atom (Mn or Fe)
and \FeAC. The \FeB-\FeAC\ interaction, $J_1(\mbox{Fe-Fe})$, is
between 20 and 10~meV, depending on concentration, and the
Mn$_{\mathrm{B}}$-\FeAC\ interaction, $J_1(\mbox{Mn-Fe})$, is between
3-5~meV; next-nearest neighbor interactions (Fe-Fe, Mn-Mn or
Fe-Mn) are typically at least one order of magnitude weaker. Therefore
we expect the Curie temperature trend with concentration to follow the
behavior of the averaged $J_1$, at least qualitatively.

\begin{figure}
  \begin{center}
    \includegraphics[width=8cm]{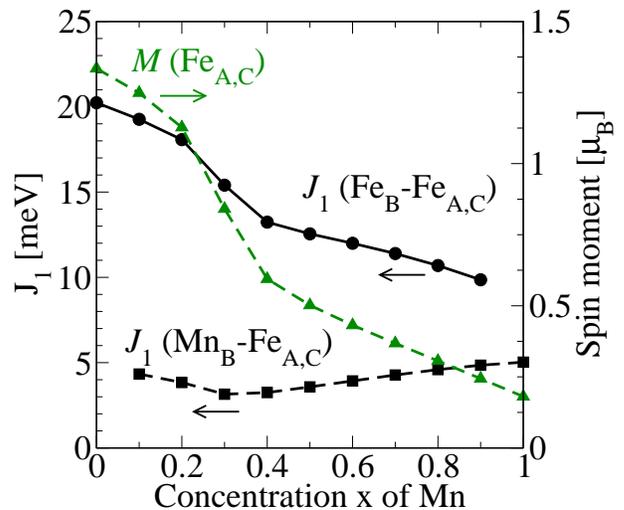}
  \end{center}
  \caption{(color online) Nearest neighbor exchange interactions $J_1$
    between \FeB\ and \FeAC\ atoms and between Mn$_{\mathrm{B}}$ and
    \FeAC\ atoms, as calculated by the Liechtenstein formula
    (\ref{eq:liechtenstein}). The \FeAC\ moments, $M$(\FeAC), are also
    shown, to demonstrate the correlation between the decrease of
    $J_1$ and $M$(\FeAC). \label{fig:J1}}
\end{figure}

The trends of $J_1$ with concentration are depicted in
Fig.~\ref{fig:J1}, together with the moment of the \FeAC\ atoms,
$M$(\FeAC). There is a clear correlation between
$J_1(\mbox{\FeB-\FeAC})$ and $M$(\FeAC), while $J_1(\mbox{Mn-\FeAC})$
seems unaffected by the drop of $M$(\FeAC). We now discuss these
observations. The pair-interaction energies $E_{ij}=-J_{ij}
\,\hat{e}_{i} \cdot \hat{e}_{j}$, determined by the electronic
structure, contain the absolute value of the atomic spin moments in a
non-trivial way. By this we mean that, if the moments are varied by
some external parameter ({\it e.g.}, here, by changing the concentration),
$E_{ij}$ can be affected either just by the variation of the absolute
value of the moment (as suggested by the alternative form derived from
Eq.~(\ref{eq:heisenbergB}),
$E_{ij}=-\tilde{J}_{ij}\,\vec{M}_i\cdot\vec{M}_j$), or also by an
alteration of the exchange mechanism, which is induced by the change
of the electronic structure via the external parameter and affects the
constants $J_{ij}$.

Apparently, in \FMS\ we are faced with both situations. On the one
hand, the dominant trend for $J_1(\mbox{Fe-Fe})$ comes from the
reduction of the \FeAC\ moment as the Mn concentration $x$ increases,
although there seems to be also an alteration of $\tilde{J}_1$, since
the drop of $M$(\FeAC) is faster than the drop of $J_1$. On the other
hand, $J_1(\mbox{Mn-Fe})$ is left practically unaltered \emph{despite}
the strong reduction of $M$(\FeAC). In order to interpret this, we
again observe the shifting of the peak, indicated by an arrow in the
\FeAC\ DOS in Fig.~\ref{fig:dos}, as a function of $x$. This
hybridization-induced peak coincides more and more with its associated
Mn peak as the Mn concentration is increased. Since the Fermi level
bisects the Mn peak (and increasingly more the \FeAC\ peak), the
double-exchange mechanism sets in progressively more and more,
favoring ferromagnetic alignment of the moments. (We remind the reader
that the double-exchange mechanism is present when half-filled states
of the same spin hybridize with each other, resulting in a band
broadening and a gain in energy; in a tight-binding picture, the
kinetic energy is lowered by the inter-atomic hopping of electrons,
allowed by the half-filled band.)  Thus, the progressive shift of the
indicated peak in the \FeAC\ DOS causes two competing effects: a
reduction of the moment $M$(\FeAC) (as discussed in
Sec.~\ref{sec:dos}) and a strengthening of the Mn-Fe pair exchange
interaction. These effects by and large cancel each other in the
Heisenberg energy expression, and the net result is only a weak
dependence of $J_1(\mbox{Mn-Fe})$ on concentration.

\begin{figure}
  \begin{center}
    \includegraphics[width=8cm]{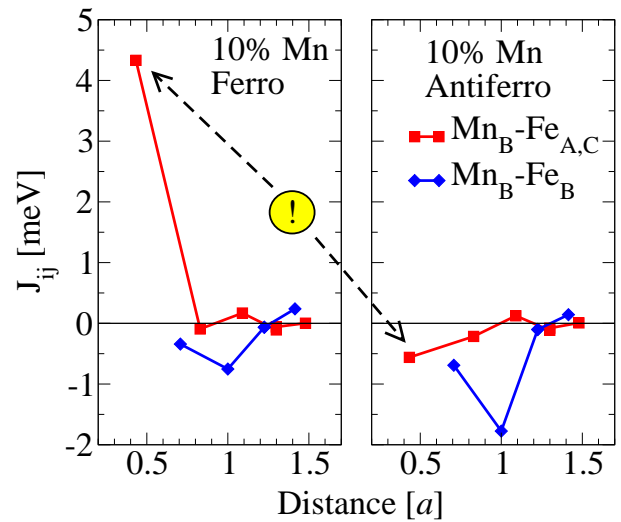}
  \end{center}
  \caption{(color online) Calculated Mn-Fe exchange interaction
    $J_{ij}$ via the Liechtenstein formula (\ref{eq:liechtenstein})
    for a Mn concentration of $x=0.1$, starting from two different
    configurations. Left: ferromagnetic starting point ({\it i.e.}, with the
    Mn moments ferromagnetically aligned to the Fe$_3$Si matrix),
    which is also the ground state. Right: antiferromagnetic starting
    point ({\it i.e.}, with the Mn moments antiferromagnetically aligned to
    the Fe$_3$Si matrix). There is a strong qualitative difference in
    the dominant nearest-neighbor interaction, emphasized by an
    exclamation mark, while the more distant interactions change
    quantitatively, but not so much qualitatively. This discrepancy is
    not observed at higher Mn concentrations, when the electronic
    structure of the \FeAC-atoms at \EF\ is dictated more by the
    hybridization of their $d$-states with the Mn $d$-states. The
    lines are guides to the eye.\label{fig:Mn10_F_AF}}
\end{figure}

We close the discussion on the exchange parameters with the following
comments on the calculations. In the present work, the exchange
constants were calculated starting from the ground-state
(ferromagnetic) configuration. As a test, however, we calculated the
Mn-Fe exchange constants starting from the antiferromagnetic (AF)
state ({\it i.e.}, with the Mn moments antiferromagnetically aligned
to the Fe$_3$Si matrix) for a Mn concentration of $x=0.1$, and
compared with the result starting from the ferromagnetic (FM) ground
state. As can be seen in Fig.~\ref{fig:Mn10_F_AF}, there is a strong
qualitative difference in the calculated value of the nearest-neighbor
Mn-Fe interaction in the two cases: starting from FM, we obtain a
tendency to retain ferromagnetism (positive exchange constant); while
starting from AF, we obtain a tendency to retain antiferromagnetism
(negative exchange constant). For the more distant neighbors the two
calculations give quantitative, but not so much qualitative,
differences. This discrepancy demonstrates the significant change in
electronic structure at high angles, far beyond the assumptions of a
Heisenberg model. The discrepancy is not observed at higher Mn
concentrations, when the electronic structure of the \FeB-atoms at
\EF\ is dictated more and more by the hybridization of their
$d$-states with the Mn $d$-states. In addition, we note that, in the
case of Fe$_2$MnSi, the LSDA result for the Mn-\FeAC\
nearest-neighbor exchange parameter is weaker by approximately $1/3$
compared to the GGA result. This reflects also on the magnon spectra
that are discussed in Sec.~\ref{sec:limitations}. Such a discrepancy
between LSDA and GGA does not show up for Fe$_3$Si.

Calculated Curie temperatures $\Tc$ of \FMS\ are shown in
Table~\ref{tab:Curie} together with experimental results. The
experimental finding of a reduction of $\Tc$ with increasing
concentration is reproduced, although the calculated results
systematically underestimate the experimental values. The reason for
this trend is obviously the reduction of $J_1(\mbox{Fe-Fe})$ as a
function of concentration, in conjunction with the comparatively low
values of $J_1(\mbox{Mn-Fe})$ which become important at high
concentrations. Note that, at $x=1$, the Fe$_2$MnSi alloy is found
experimentally\cite{Ziebeck76} to possess a degree of disorder in the
form of a Mn$_{\mathrm{B}}$-\FeAC\ swap of 12\%. When considering such
a swap in the calculations, we found an increased $\Tc$ of 200~K,
mainly because the Fe atoms replacing Mn at the B-site have a stronger
exchange interaction with the \FeAC\ neighbors. However, a possible Mn
clustering at this concentrations, which could affect the value of the
exchange interactions, cannot be taken into account within the
CPA. This is discussed in more detail in the next section.

\begin{table}
  \caption{Calculated Curie temperature of \FMS\ at $x=0$, 0.5, and
    1. Experimental results (Ref.~\onlinecite{Yoon77}) are also
    shown.\label{tab:Curie}}
  \begin{tabular}{lcl}
    \hline
    $x$ \phantom{space} & $\Tc$ [K] (exp.) & $\Tc$ [K] (calculated) \\
    \hline
    0.0 & 803          & 730 \\
    0.5 & 450          & 320 \\
    1.0 & 200          & 160 (ordered; Mn at B) \\
    &              & 200 (12\% Mn at A,C)\\
    \hline
  \end{tabular}
\end{table}

\begin{figure}
  \begin{center}
    \includegraphics[width=8cm]{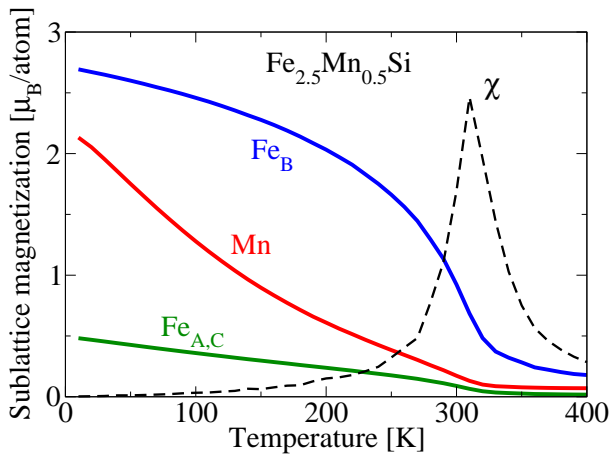}
  \end{center}
  \caption{(color online) Monte Carlo simulation of the magnetization
    and susceptibility $\chi$ {\it vs.} temperature for
    Fe$_{2.5}$Mn$_{0.5}$Si. It is evident that the Mn sublattice
    magnetization drops much faster than the Fe magnetization. The
    susceptibility peak signals the Curie
    temperature.\label{fig:MvsT}}
\end{figure}

Before closing this section, we show an interesting behavior of the
Mn sublattice magnetization at low or intermediate concentrations. Kepa {\it et
al}.\cite{Kepa88} and Ersetz {\it et al}.\cite{Ersez95} find experimentally
that the \MnB\ moment at room temperature is significantly lower than
what is reported close to $T=0$~K. {\it E.g.}, for $x=0.55$,\cite{Ersez95}
$M({\rm Mn})=1.42$~\mub\ at room temperature instead of
2.2~\mub. Therefore they propose that the Mn average moment drops with
temperature faster than the Fe moment. This is reproduced by our Monte
Carlo simulations, and is due to the weak coupling of Mn to \FeAC\
compared to the coupling of \FeB\ to \FeAC.  As we see in
Fig.~\ref{fig:MvsT}, the Mn magnetization curve does not follow the
critical-behavior form with an inflection point of the Fe curve, but
rather drops almost linearly. This is also reflected in the Mn
sublattice susceptibility (not shown here), which does not show a
peak at $T=\Tc$.

\section{Discussion on the re-entrant behavior; Limitations of present
  calculations\label{sec:limitations}}

In Sec.~\ref{sec:background} we summarized what is known on the
anomalous behavior of the magnetization of \FMS\ for $x>0.75$. The
magnetization $M(T)$ increases for $0<T<\TR$ (contrary to the behavior
in a usual ferromagnet), then decreases again up to $\Tc$; $\TR$
varies with concentration, starting from low values at $x=0.75$ and
saturating at about 70-80~K at $x=1$. This so-called \emph{re-entrant}
behavior and the resulting \emph{re-ordered} phase, which has been the
focus of many experimental works, could not be reproduced by
calculations within the approximations used in the present work.

There is experimental evidence,\cite{Ziebeck76,Yoon77} based on
neutron scattering data, that the re-entrant behavior arises from
transverse ordering of the Mn$_{\rm B}$ magnetic moments. This means that,
in the ground state, the Mn moments are partly canted to the direction
of average magnetization. As the temperature is increased, the Mn
moments absorb energy by aligning their spins and thus the average
magnetization increases.

Mohn and Supanetz\cite{Mohn98} performed first-principles (LSDA-based)
calculations of non-collinear magnetic structures for spin-spirals of
several wavevectors $\vc{q}$. For ordered Fe$_2$MnSi they found that
the ferromagnetic state represents a local minimum, while a spin
spiral of wavevector $\vc{q}=(2\pi/a)(0.45,0.45,0.45)$ has a lower energy
by about 0.75~mRyd (10.2~meV). The absolute energy minimum which
they found is very slightly lower, for a state which shows a canted
antiferromagnetic ordering of the Mn moments along the [111] axis, with
a canting angle of $60^\circ$ with respect to the [111] axis.

\begin{figure}
  \begin{center}
    \includegraphics[width=6cm]{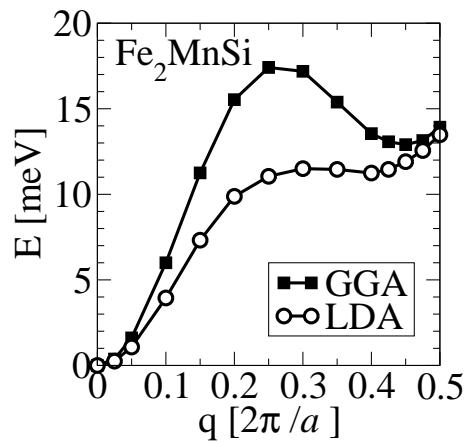}
  \end{center}
  \caption{(color online) Calculated flat spin-spiral dispersion
    relations of Fe$_2$MnSi along the [111] direction within the LSDA
    and GGA. The abscissa $q$ corresponds to the point $(q,q,q)2\pi/a$
    of the Brillouin zone. The FLAPW method was used in this calculation.\label{fig:magnons}}
\end{figure}

We also performed calculations of magnetically non-collinear spin
spirals along the [111] axis using the FLAPW method.\cite{FLEUR}
However, the results of Ref.~\onlinecite{Mohn98} were not
reproduced. Our results are shown in
Fig.~\ref{fig:magnons}. Calculations within both the LSDA and the GGA
show a ferromagnetic ground state ({\it i.e.}, a global minimum at
$\vc{q}=0$), while a local minimum is found close to
$\vc{q}=(2\pi/a)(0.45,0.45,0.45)$, {\it i.e.}, at the point where the
spin-spiral minimum is found in Ref.~\onlinecite{Mohn98}. The energy
difference between the two (local and global) minima is of the order
of 11~meV within the LSDA and 13~meV within the GGA. We do not know
the origin of the discrepancy between our calculations and the
calculations of Ref.~\onlinecite{Mohn98}. A possible source of
discrepancy is the use of a spherical potential approximation in
Ref.~\onlinecite{Mohn98}, as opposed to a full potential calculation
here.

However, these non-collinear calculations (ours as well as the ones of
Ref.~\onlinecite{Mohn98}) neglect a 12\% Mn-Fe swap which is seen
experimentally in Fe$_2$MnSi; {\it i.e.}, experimentally this alloy has a
small degree of disorder. Given the small calculated energy
differences between the local and global minima (about 10~meV per
formula unit), a correct description of this swap can have important
consequences. We attempted to model the swap by calculating the
electronic structure of
(Fe$_{0.94}$Mn$_{0.06}$)$_{\mathrm{A,C}}$(Fe$_{0.12}$Mn$_{0.88}$)$_{\mathrm{B}}$Si
within the CPA; here, the indexes (A,B,C) refer to the corresponding
positions in the unit cell. The magnetic structure was subsequently
investigated both by calculating the exchange constants and the
resulting configuration at $T=0$~K and by calculating a possible
(collinear) disordered local moment state at the central
site. However, in all cases the outcome was a ferromagnetic ground
state.

What is missing within the CPA description is the short-range
configurational order, {\it i.e.}, the possibility to describe clustering of
Mn$_{\mathrm{A,C}}$ atoms around Mn$_{\mathrm{B}}$ atoms. We speculate
that this swap and clustering causes the canting of some
Mn$_{\mathrm{B}}$ moments by interaction with the neighboring
Mn$_{\mathrm{A,C}}$ atoms and is therefore essential for the
appearance of the re-entrant behavior (such a scenario was already
suggested in Ref.~\onlinecite{Niculescu76}). Possibly, as the
temperature is increased, the canting of the Mn$_{\mathrm{B}}$ moments
is reduced on the average and the total magnetization increases; this
hypothesis would require a weak coupling of the canted
Mn$_{\mathrm{B}}$ moments to their Mn$_{\mathrm{A,C}}$ neighbors
compared to the coupling of the non-canted Mn$_{\mathrm{B}}$ moments
to their \FeAC\ neighbors.

Our hypothesis is supported by the following experimental
findings. (i) The re-entrant behavior and the swap appear in the same
concentration range ($x>0.75$). (ii) The re-entrant behavior is
sensitive to the heat treatment of the alloy; after proper annealing,
it was found that the re-entrant behavior smoothens, although $\TR$
does not change.\cite{Yoon77} This, in conjunction with the
calculations showing that the un-swapped state is the ground state,
suggests that annealing causes a fraction of the Mn atoms to
return from the A,C site to the B-site, so that the number of Mn
clusters (and canted Mn moments, as we suspect) lessens. If this
hypothesis is true, $\TR$ should be indeed unaffected by annealing,
because it would correspond to a finite-size effect (characteristic
exchange-energy scale of a small cluster) rather than a phase
transition. Such a possibility has been suggested by Nielsen and
collaborators,\cite{Nielsen96} based on Monte Carlo calculations of
model systems. (iii) Clustering would result in a local environment
which is closer to Mn$_3$Si, which is known to show antiferromagnetic
behavior.

Furthermore, it is observed that the ``smoothening'' of the re-entrant
behavior after annealing is accompanied by an increase of both $M_0$
and $M_{\rm extr}$ (shown in Fig.~7 of Ref.~\onlinecite{Yoon77} for
Fe$_{1.75}$Mn$_{1.25}$Si). We therefore make the plausible assumption
that this behavior is present also at concentrations $x<0.75<1$, and
that the ground state, with all Mn atoms being at the B-site, will
show a higher magnetic moment than $M_{\rm extr}$. This could resolve
the discrepancy between the measured\cite{Yoon77} magnetization value
of $M_{\rm extr}\approx 2.3~\mub$ compared to calculated value of 3~\mub\
in Fe$_2$MnSi.

First-principles investigations of such short-range-order effects
require an approach beyond the CPA, {\it e.g.} by use of the non-local
CPA\cite{NLCPA} or large-supercell techniques,\cite{KKRnano} and we
therefore defer this study to a future work.

\section{Conclusion\label{sec:conclusion}}

The electronic and magnetic structure of the magnetic intermetallic
alloy \FMS, for $0 \leq x\leq 1$ has been investigated using
density-functional theory within the GGA, together with the CPA to
describe disorder. We find that important experimental findings, such
as the trends of the magnetization and of the Curie temperature as a
function of concentration, are reproduced. They can be interpreted in
terms of single-particle energies with the help of the density of
states, by using simple physical arguments, namely symmetry-dependent
hybridization of wavefunctions and local charge neutrality.
Quantitatively, the drop of the magnetization as a function of Mn
concentration is underestimated, especially at $x>0.75$, however this
could be due to Mn-Fe swap and Mn clustering in experiment that cannot
be captured by the CPA; the same applies to the re-entrant
behavior. It is therefore worthwhile to investigate \FMS\ beyond this
approximation in the future.

However, it is rather surprising that the two most common
approximations to DFT, namely the LSDA and the GGA, give ground states
that differ at low concentration not only quantitatively but also
qualitatively, although they almost agree on the single-particle
spectrum. The superiority of GGA that we find here is common but
certainly not general, which shows the need for theory to go
hand-in-hand with experiment for the understanding of magnetic
intermetallic compounds.

Concerning the relevance to spintronics, the most important finding
here is the continuous variation of the spin polarization at \EF\ over
a wide range, between $-0.3<P<1$ for $0<x<0.75$, {\it i.e.}, in the
region where theory and experiment are in reasonable agreement and
before the onset of the re-entrant behavior. Since the polarization at
\EF\ is a property that is notoriously difficult to measure with
precision, the variation which is found theoretically could be used to
improve or calibrate the methods of measurement of $P$ in a single
type of material, so that spurious effects in measurement can be
treated on the same footing and understood better.

\end{document}